\documentclass[fdp,a4paper,fleqn]{w-art}
\usepackage{hyperref}
\usepackage{microtype}
\usepackage[english]{babel}
\usepackage{times}
\usepackage{cite}
\usepackage{amsmath, amsfonts, amssymb}
\usepackage{bbm}
\usepackage[]{graphicx}

\numberwithin{equation}{section}

\DeclareMathOperator{\re}{Re}
\DeclareMathOperator{\im}{Im}

\renewcommand{\Im}{\im}

\DeclareMathOperator{\Tr}{Tr}

\DeclareMathOperator{\Vol}{Vol}

\newcommand{\abs}[1]{\left\lvert#1\right\rvert}

\newcommand{\cC}{{\mathcal{C}}}

\newcommand{\cL}{{\mathcal{L}}}

\newcommand{\cN}{{\mathcal{N}}}

\newcommand{\cZ}{{\mathcal{Z}}}

\newcommand{\bbR}{\mathbb{R}}
\newcommand{\bbC}{\mathbb{C}}
\newcommand{\bbZ}{\mathbb{Z}}

\begin{document}


\title{Unoriented D-brane instantons}

\author[M. Bianchi]{Massimo Bianchi\inst{1}%
	\footnote{%
		Corresponding author\quad E-mail:~\textsf{Massimo.Bianchi@roma2.infn.it},
		Phone: +39\,06\,7259\,4559,
		Fax:   +39\,06\,202\,5259
	}
}
\author[G. Inverso]{Gianluca Inverso\inst{1}}
\address[\inst{1}]{%
	Dipartimento di Fisica, Universit\`a di Roma Tor Vergata\\
	and I.N.F.N. - sezione di Roma 2,
	via della Ricerca Scientifica, 00133 Roma, Italy
}

\begin{abstract}
We give a pedagogical introduction to D-brane instanton effects in vacuum configurations with open and unoriented strings. We focus on quiver gauge theories for unoriented D-branes at orbifold singularities and describe in some detail the $\bbZ_3$ case, where both `gauge' and `exotic' instantons can generate non-perturbative super potentials, and the $\bbZ_5$ case, where supersymmetry breaking may arise from the combined effect of  `gauge' instantons and  a FI D-term. 
\end{abstract}

\maketitle

\vspace*{-1.1em}
\section{Introduction}

String theory is widely believed to offer the possibility of unifying gravity and the other interactions at the quantum level
\cite{Kirbook}. In modern approaches, gravity is mediated by closed strings while gauge interactions are mediated by open strings, that represent the low-energy excitations of D-branes \cite{stringpheno}. Scenarios with large extra dimensions and low scale string tension are viable but many problems related to supersymmetry breaking and moduli stabilization are still open and call for the introduction of internal fluxes and non-perturbative effects associated to Euclidean branes \cite{BCKWreview}.

Aim of this lecture is to give a pedagogical introduction to D-brane instanton effects in vacuum configurations with open and unoriented strings. We will start with the ADHM construction of Yang-Mills instantons \cite{ADHM} and describe how it finds an intuitive almost natural embedding in open string theories \cite{Douglas'95, Billo, MKR}. We then pass to consider the two broad classes of D-brane instantons: `gauge' instantons and `exotic' instantons \cite{Billoetal2009, FMP}. The former correspond to ED-branes wrapping the same cycle as a physical stack of D-branes. The latter correspond to ED-branes wrapping a cycle not wrapped by any physical stack of D-branes.
For definiteness we will start with the simplest case: ED(-1)-branes (or `D-instantons') on a stack of D3-branes.
We then pass to consider quiver gauge theories governing the low-energy dynamics of (unoriented) D-branes at an orbifold singularity. We will review in some detail the case $\bbR^6/\bbZ_3$ with both unoriented $\Omega$-projections and sketch the derivation of the non-perturbative super potentials of the VY-ADS type (generated by `gauge instantons') and of the `exotic' type (generated by `stringy' instantons) \cite{U(4)}. We highlight the role of the super-moduli and of the anomalous U(1) symmetries. We also briefly address other cases with different amount of supersymmetry and/or different structure of fermonic zero-modes that determine the nature of the instanton effect (superpotential \cite{Blum, U(4)}, threshold correction to gauge kinetic function \cite{ABDFPT, Camara:2007dy}, higher-derivative term \cite{BWnieist}, etc.). Finally we discuss a toy model based on the $\bbR^6/\bbZ_5$ quiver theory with gauge group $U(1)\times U(5)$ where dynamical supersymmetry breaking is triggered by the combined effects of `gauge' instantons and a non-vanishing FI term for the first $U(1)$  \cite{MBJFMFF}.
We conclude with an outlook and directions for future investigation.

\section{Instantons in Gauge and String Theory}
In Yang--Mills theory instantons are solutions to the equations of motion with finite Euclidean action. They generate non-perturbative effects with  strength $e^{-1/g^2_{YM}}$, $g_{YM}$ being the gauge coupling constant.
They saturate a BPS-like bound and satisfy an (anti) self-duality condition (in Euclidean signature):
\begin{equation}
 \tilde F_{\mu\nu} = \frac{1}{2}\epsilon_{\mu\nu\rho\sigma}F^{\rho\sigma} = \pm F_{\mu\nu}.
\end{equation}
Defining the complexified gauge coupling $\tau = \frac{\theta}{2\pi} + i \frac{4\pi}{g^2_{YM}}$, the action reads
\begin{equation}
\begin{aligned}
S_{YM} = &-\frac{\im \tau}{8\pi}\int d^4x \Tr F_{\mu\nu}F^{\mu\nu}+i\frac{\re \tau}{8\pi}\int d^4x \Tr F_{\mu\nu}\tilde F^{\mu\nu} \quad .
\end{aligned}
\end{equation}
For an instanton configuration with topological charge $K$, one has $S_K = \frac{8\pi^2}{g^2_{YM}}\abs K + i \theta K $, where
\begin{equation}
\begin{aligned}
K = &-\frac{1}{16\pi^2}\int d^4x \Tr F_{\mu\nu}\tilde F^{\mu\nu} \quad .
\end{aligned}
\end{equation}

An elegant and exhaustive algebraic construction of Yang--Mills instantons is due to Atyah, Drinfeld, Hitchin and Manin (ADHM) \cite{ADHM}. For $SU(N)$ Yang-Mills theory and $K$ instantons one starts with the ansatz
\begin{equation}
A_\mu = U^\dagger \partial_\mu U
\end{equation}
with $U$ a $(2K+N)\times N$ complex matrix that satisfies $U^\dagger U = 1_{N\times N}$. The ADHM data are coded in a
$N\times 2K$ matrix $\Delta$ satisfying $U^\dagger \Delta= 0$ (i.e. spanning the kernel of $U^\dagger$). It can be shown that $\Delta$ is linear in $x$ (space-time coordinates) and can be decomposed in terms of two $K\times N$ matrices $w_{\dot\alpha}$ and four $K\times K$ matrices $a_\mu$ according to $\Delta = w \oplus (a + x)$.
In order to ensure self-duality of the field strengths one must impose the ADHM constraint ($q=1,2,3$):
\begin{equation}\label{adhm constraint}	
	w_{\dot\alpha\,u}{}^a (\sigma^q)^{\dot\alpha}{}_{\dot\beta} \bar w^{\dot\beta}{}_b{}^u -i \bar\eta^q_{\mu\nu} [a^\mu ,a^\nu]^a{}_b = 0 \quad ,
\end{equation}
$\eta^q_{\mu\nu},\ \bar\eta^q_{\mu\nu} $ are the 't Hooft symbols and
$(\sigma^q)^{\dot\alpha}{}_{\dot\beta} = \frac{i}{2} \bar \eta^q_{\mu\nu} (\bar\sigma^{\mu\nu})^{\dot\alpha}{}_{\dot\beta}$, with $\sigma^\mu = (\mathbbm 1, \sigma^q)$ in Euclidean signature.
Equation \eqref{adhm constraint} has a $U(K)$ invariance. Once this has been taken into account, the moduli space $\mathfrak M_K$ of instantons with topological charge $K$ turns out to be a $4KN$ dimensional HyperK\"ahler manifold.
Indeed, we have $3K^2$ equations for $4KN + 4K^2$ variables, out of which $K^2$ are redundant due to $U(K)$ invariance.
The moduli $a_\mu ,\ w_{\dot\alpha},\ \bar w^{\dot \alpha}$ parametrize the position, size and gauge orientation of the  instanton. As we will momentarily see, they can be nicely interpreted as the bosonic massless modes of open strings in a D$p$ ED$(p-4)$ brane system. Moreover $U(K)$ invariance is interpreted as the symmetry arising from the Chan--Paton factors of strings ending on the Euclidean branes.

\subsection{Instantons from branes}

It is remarkable how the ADHM construction \cite{ADHM} admit a nice geometrical interpretation in string theory, in terms of systems of D-branes \cite{Douglas'95}. We will therefore display how and why one should expect gauge instantons in (super) Yang--Mills theory to be implemented in String Theory as systems of D-branes.
Let us start with the (Euclideanized) DBI world-volume action of a D$p$-brane wrapping a $(p-3)$ cycle $\mathcal C$ in internal space, we write:
\begin{align}\notag
	S = & \mu_p \Tr \left[ \int_{\mathbb R^4\times \mathcal C}
	d^4x\ e^{-\varphi} \sqrt{\det(g+2\pi\alpha' F)} \right]
	-i  \int_{\mathbb R^4\times \mathcal C} \sum_n C_{2n} e^{2\pi\alpha'F}
\end{align}
Here $\mu_p = g_s(2\pi)^{-p} (\alpha')^{-(p+1)/2}$ is the D$p$ brane tension.
We have also assumed that the bulk field $B_{\mu\nu}$ vanish.
Expanding at the quadratic level in $F$ one sees that the complex gauge coupling $\tau$ is given by the expression:
\begin{equation}
 2\pi \, \tau = \mu_{p-4} i \int_{\mathcal C} e^{-\varphi}\sqrt{g} + C_{p-3}
\end{equation}
which reads exactly as the contribution from an ED$(p-4)$ brane wrapping $\mathcal C$ and matches the contribution to the partition function of a $K=1$ instanton.
Indeed, the contribution from a localized source of $C_{p-3}$ within the D$p$ brane would read $\sim \int C_{p-3}\wedge\Tr( F\wedge F )$, hence sourcing an instanton contribution.
This suggests that gauge instantons can be described as bound states of $N$ D$p$ branes and $K$ ED$(p-4)$ branes wrapping the same cycle in internal space.
If the two cycles are different, one obtains a contribution from the Euclidean branes wrapping a cycle $\mathcal C'$ of the form $2\pi\,\tau'$, with $\tau'$ not directly related to the gauge coupling. This are `stringy' or `exotic' instantons, whose importance will be discussed later on.

\subsection{The D3 D$(-1)$ system}

A good starting point to discuss gauge instantons in String Theory is the D3 D$(-1)$ system, corresponding to $\mathcal N=4,\ D=4$  $U(N)$ super Yang--Mills theory.
Open strings stretching between D3 branes give rise to the dynamical fields of the $\mathcal N=4$ vector supermultiplet.

Strings with at least one end on the D$(-1)$ branes have no dynamical massless modes.
D$(-1)$ D$(-1)$ strings give rise to a matrix model which can be seen as obtained from dimensional reduction to $D=0$ of $\mathcal N=1$ SYM theory in 10 dimensions with $U(K)$ gauge group. The corresponding moduli are
\begin{equation}
 a_\mu{}_a{}^b,\quad \chi_i{}_a{}^b,\quad \Theta^A_\alpha{}_a{}^b,\quad \bar \Theta_A^{\dot\alpha}{}_a{}^b.
\end{equation}
The first two are the bosons decomposed in one (nondynamical) vector in the longitudinal directions of the D3 brane and six scalars ($i=1,\ldots,6$), while $\Theta ,\bar \Theta$ are the superpartners, $A=1,\ldots,4$ being a Weyl index with respect to $SO(6)$.
We will also introduce the three off-shell auxiliary fields $D^q,\ q=1,\ldots,3$.
In particular, $a_\mu{}_a{}^b$ and $\chi_i{}_a{}^b $ parametrize the positions of the D$(-1)$ branes (and hence of the gauge instanton) in the directions longitudinal and transverse to the D3 brane, respectively.
Similarly, D3 D$(-1)$ strings give rise to moduli in the $\bf N\bar K$ and $\bf \bar N K$ representations of $U(N)\times U(K)$, hence charged under the gauge group of the SYM theory. They are:
\begin{equation}
 w_{\dot\alpha\,u}{}^a,\  \nu_{A\,u}{}^a,\quad
 \bar w^{\dot\alpha\,u}{}_a,\  \bar\nu^{A\,u}{}_a.
\end{equation}
In this case $w_{\dot\alpha},\ \bar w^{\dot\alpha} $, which are independent of each other, are bosons and parametrize the size and orientation of the instanton, while $\nu ,\ \bar\nu$ are their super-partners.

In order to show that $a,w,\bar w$ are indeed the ADHM parameters, one should now reconstruct their equations of motion from string theory.
It can be shown that string amplitudes on the disk, with the insertion of only the non-dynamical ADHM moduli and their superpartners, reproduce their action.
Starting with the D$(-1)$ D$(-1)$ string, the vertex operators (in the $-1$ picture) for the bosons read:
\begin{align}
 V_{a} = a_\mu e^{-\varphi} \psi^\mu T_{\bf K\bar K},\qquad
 V_{\chi} = \chi_i e^{-\varphi} \psi^i T_{\bf K\bar K},
\end{align}
where $\psi^\mu$ are the worldsheet fermions, $e^{-\varphi}$ denotes the bosonization of the worldsheet superghosts and $T$ is the $U(K)$ Chan-Paton matrix. One can similarly define the vertex operators for the superpartners $\Theta ,\ \bar\Theta$ by introducing a spin field $S_r$, $r=1,\ldots,16$ (which keeps into account both $SO(4)$ and $SO(6)$ Weyl indices and both chiralities):
\begin{equation}
 V_{\Theta} = \Theta^r S_r e^{-\varphi/2} T_{\bf K\bar K}
\end{equation}
Spin fields are worldsheet operators that have branch-cut OPE with worldsheet fermions $\psi^\mu$ and allow to modify their mode expansion and monodromy, effectively switching between the NS and R sectors.
Strings in the D3 D$(-1)$ sector have mixed Neumann--Dirichlet (N-D) boundary conditions along four directions. To account for this, vertex operators involve twist fields $\sigma_\mu$, that play for worldsheet bosons a similar role to spin fields for fermions\footnote{Twist fields appear for mixed boundary conditions and, more in general, configurations with branes at angles \cite{Bertolini:2005qh}}. The relevant expressions read
\begin{align}
 V_{w} &= \sqrt{\frac{g_s}{v_{p-3}}} w_{\dot\alpha} e^{-\varphi} \prod_\mu \sigma_\mu C^{\dot\alpha} T_{\bf N\bar K} ,\qquad
 V_{\nu} = \sqrt{\frac{g_s}{v_{p-3}}} \nu^A e^{-\varphi/2} \prod_\mu \sigma_\mu C_A T_{\bf N\bar K}
\end{align}
Again, $C^{\dot\alpha},\ C^A$ are spin fields, respectively in the spinor representations of $Spin(4)$ and $Spin(6)$.
The normalization is crucial in order to recover the correct field theory limit for $\alpha'\rightarrow0$.

The result for the ``action'' of the instanton moduli reads schematically:
\begin{equation}
 S_{N,K} = \Tr \left(\frac{1}{g^2_0}S_G + S_K + S_D\right),
\end{equation}
Where $g^2_0 = {g_s}/{(4\pi^3\alpha'^2)}$. The trace is over $U(K)$ indices and the various terms are:
\begin{align}
	S_G = &-[\chi_i,\chi_j]^2 +i\bar\Theta_{\dot\alpha\,A}[\chi^\dagger_{AB},\bar\Theta^{\dot\alpha}_B] -D^qD^q															 \\
	 S_K\, = &-[\chi_i,a_\mu]^2 +\chi^i\bar w^{\dot\alpha} w_{\dot\alpha} \chi_i 
	-i \Theta^{\alpha A}[\chi_{AB},\Theta_\alpha^B] +2i \chi_{AB}\, \bar\nu^A\nu^B	\\[.3em]
	 S_D = &\ i\left(-[a_{\alpha\dot\alpha},\Theta^{\alpha A}]+\bar\nu^A w_{\dot\alpha} + \bar w^{\dot\alpha} \nu^A\right) \bar\Theta_A^{\dot\alpha}
					+ D^q \left(\bar w \sigma^q w -i\bar\eta^q_{\mu\nu}[a^\mu ,a^\nu]\right).
\end{align}
The 6-dimensional chiral ``sigma'' matrices $\Sigma^i_{AB}$ are used to write $\chi_{AB} = \frac{1}{2}\chi_i \Sigma^i_{AB}$, and they can be chosen to be written in terms of the `t Hooft symbols as $\Sigma^i_{AB} =(\eta^q_{AB}, \bar \eta^{q+3}_{AB}) $.
The first term decouples in the limit $\alpha'\rightarrow0$ (while keeping $g_s$ fixed) and $D^q$ become Lagrange multipliers. The resulting equations of motion are the ADHM constraint \eqref{adhm constraint} and its supersymmetric counterpart:
\begin{equation}
 	[a_\mu ,\Theta^A]\sigma^\mu + w\bar\nu^A + \nu^A\bar w = 0.
\end{equation}
Following the same lines, string amplitudes on the disk with the insertion of dynamical fields provide the (super)instanton profiles and their non-perturbative contributions to scattering amplitudes.

An extra term arises in order to take into account vev's for the dynamical scalar fields $\varphi^i$of the D3 D3 sector.
In that case one adds the couplings
\begin{equation}\label{S phi}
	S_\varphi = \Tr\left( \bar w^{\dot\alpha}(\varphi^i\varphi_i + 2\chi^i\varphi_i)w_{\dot\alpha} + 2i \bar \nu^A \varphi_{AB} \nu^B \right).
\end{equation}
This term plays an essential role when we want to calculate the instanton contributions to the effective action.
Indeed, the instanton partition function reads:
\begin{equation}\label{instanton partition function}
\begin{split}
	\cZ_{N,K} &= \int_{\mathfrak M_K} e^{-S_{N,K} - S_\varphi}	\\
		&= \frac{1}{\Vol U(K)}
			\int d^{4K^2}\hspace{-4pt}a\ d^{6K^2}\hspace{-4pt}\chi\ d^{8K^2}\hspace{-3pt}\Theta\
			d^{8K^2}\hspace{-3pt}\bar\Theta\ d^{3K^2}\hspace{-3pt}D\ d^{4KN}\hspace{-2pt}w\ d^{8KN}\hspace{-2pt}\nu
			\ \ e^{-S_{N,K} - S_\varphi}.
\end{split}
\end{equation}

\subsection{Non-perturbative superpotentials}

Depending of the number of supersymmetries and the structure of the fermionic zero-modes, instantons can generate different kinds of non-perturbative effects. For instance, in  ${\cal N}=4$ theories in $D=4$ they contribute to correlation functions of more than three CPO operators, while in ${\cal N}=2$ theories in $D=4$ they correct the holomorphic pre-potential.

In ${\cal N}=1$ theories in $D=4$, non-perturbative corrections to the superpotential may arise, provided the integration \eqref{instanton partition function} over instanton moduli space yield the chiral superspace measure.
The superpotential is obtained from the partition function, keeping into account one-loop contributions so that:
\begin{equation}
 W_{np}  = \Lambda^{\beta K} \int_{\mathfrak M} e^{-S_{K,N}-S_\varphi}
\end{equation}
with the prefactor arising from tree and one-loop string amplitudes (respectively on the disk and on the annulus and M\"obius strip in an unoriented theory) with at least one end on the D$(-1)$'s. It takes the form
\begin{equation}
 \Lambda^{\beta K} = e^{2\pi i K \tau(\mu)},\qquad \tau(\mu)=\tau - \frac{\beta}{2\pi i}\ln\frac{\mu}{\mu_0},	
\end{equation}
where $\beta$ is the one-loop beta function coefficient. This factor compensates the scaling of the moduli space integral.
A superpotential is generated if and only if the integral over the instanton moduli space reduces to a chiral integral over $\cN=1$ superspace, with $a^\mu\rightarrow x^\mu$ and the Grassman variable $\theta_\alpha$ coming from $\Theta^A_\alpha$.
This means that we need \textit{two} exact (unlifted) fermionic zero-modes $\theta_\alpha$.
For `gauge' instanton, integration over the bosonic moduli $w$ and $\bar w$ yields negative powers of the dynamical fields and the superpotential takes the form of Veneziano--Yankielowicz/Affleck--Dine--Seiberg superpotential\begin{equation}\label{ADS}
{W_{\text{VY-ADS}}} \sim\int d^4\hspace{-1pt}x\ d^2\hspace{-1pt}\theta \frac{\Lambda^{\beta K}}{\Phi^{\beta K -3}},
\end{equation}
where $\Phi$ is the chiral superfield associated to $\varphi$. Note that the power of $\Phi$ is completely fixed by dimensional analysis.

The counting of the fermionic zero modes can be done relying on index theorems. In order to have a superpotential contribution the difference between the number of gaugino zero-modes and matter fermion zero-modes must be two: $n(\lambda_0) - n(\psi_0)= 2KN - K\sum_{\bf R} \ell_2 ({\bf R}) = 2$ , where $\ell_2({\bf R})$ is the Dynkin index of the representation $\bf R$.

\subsection{``Exotic'' Instantons}

As previously mentioned, Euclidean D$p'$ branes wrapping cycles $\cC'$ different from the ones wrapped by the physical D$p$ branes source effects that scale as $e^{-V_{p'+1}(\cC') / g_s l_s^{p'+1}} \neq e^{-1/g^2_{YM}}$, .
The prototype configuration is the ED$1$ D9 system, with 8 mixed Neumann--Dirichlet directions (as opposed to the 4 mixed directions in the ED$(p-4)$ D$p$ case).
As they scale differently than standard gauge instantons, they also source field configurations that do not satisfy self-duality conditions.
They may rather admit a field theory description in terms of octonionic instantons or hyper-instantons with $F\wedge F = \ast_8  F\wedge F$.
These exotic or stringy non-perturbative effects can provide contributions to the effective action that can neither be generated at the perturbative level nor by standard gauge instantons.
In particular, the `thumb rule' for instantons to generate corrections to the superpotential is the presence of precisely two exact unlifted fermionic zero-modes needed to produce the chiral superspace measure.
Index theorems are useless in this case because `exotic' instantons do not admit a simple field theory description.
As they arise from branes wrapping cycles different from the ones wrapped by physical branes, stringy instantons generally lack bosonic zero-modes in the charged sector (such as $w$ and $\bar w$). As a result, when they contribute to the superpotential they give rise to positive powers of the fields, produced by integration over the charged Grassmannian moduli.
`Exotic' instanton contributions will be of the schematic form:
\begin{equation}
	W_{\mathrm{exotic}} \sim M_s^{3-n} e^{-V(\cC')} \Phi^n,\qquad (n=1,2,\ldots).
\end{equation}
Such terms clearly display a very different behaviour with respect to standard instantons, as they grow for large vev's of the scalars. This would seem contradictory for gauge instantons, as asymptotic freedom suggests that these contributions should damp out for large values of the fields.
However we already stressed that the scale of exotic instantons is not directly related to $e^{-1/g^2_{YM}}$, so that a different behaviour is acceptable.

Examples of phenomenological interest where stringy instantons yield otherwise inaccessible contributions and couplings are: top-mass Yukawa coupling  in U(5) (supersymmetric) grand unified theories,
the MSSM $\mu-$term,  right-handed (s)neutrino mass terms and other that are forbidden in string perturbation theory by  anomalous U(1)'s \cite{Pascal,Lionettoetal}.
Combining the two kinds non-perturbative superpotentials, ADS-like and exotic, one can achieve (partial) moduli stabilization and supersymmetry breaking. Moreover, exotic contributions can help relaxing the tension between moduli stabilization and chirality \cite{Blumenhagen:2007sm}.

When extra zero-modes are present (e.g. for different values of $K$), the instanton will not contribute to the superpotential, but it could lead to threshold corrections to gauge kinetic functions or higher-derivative terms.
One should not forget that the counting is influenced by the presence of fluxes, both in the open and in the closed string sectors, which can lift some zero modes and hence change the nature of the effect generated by a D-brane instanton  \cite{Bianchi:2011qh, instflux, KPT}.

\section{D-branes at orbifold singularities}

A wide and interesting class of gauge theories is obtained when considering stacks of D branes at an orbifold singularity. These models combined with an unoriented projection yield indeed a promising set of theories with interesting phenomenological applications, which also allow non-perturbative effects to be computed.
We consider the case of a $\mathbb R^d/\Gamma$ singularity, with $\Gamma$ a discrete subgroup of $SO(d)$. We will focus on $\Gamma=\mathbb Z_n$ for simplicity. Putting a stack of $N$ D branes sitting at the orbifold point, these branes must be reorganized in stacks of $N_i$ ``fractional'' branes, so that $N=\sum_i N_i$ with  $i=0,\ldots,n-1$, reflecting the irreducible representations of the orbifold group.
Correspondingly, the gauge group $U(N)$ of the parent theory decomposes into $\prod_i U(N_i)$, reflecting open strings whose ends belong to the same stack and thus the same/conjugate representation.
Matter is realized in terms of open strings ending on different stacks and gives rise to massless fields in the bi-fundamentals $\bf N_i\bar N_j$.

More precisely, the $\mathbb Z_n$ orbifold projection identifies the (complex) coordinates $z^I$ according to
\begin{equation}\label{orientifold action on internal coords}
 z^I \simeq \omega^{k_I} z^I,\qquad \omega=e^{2\pi i/n}.
\end{equation}
{Preserving supersymmetry generically requires $\sum_I k_I = 0 \mod n$.}
The D branes at the singularity must therefore be regrouped so that each one has images under the action of $\mathbb Z_n$. The orbifold projection is embedded in the Chan--Paton group as
\begin{equation}\label{orientifold action on CP group}
	\rho(\mathbb Z_n) = \mathbbm1_{N_0}\oplus\,\omega\mathbbm1_{N_1}\oplus\,\omega^2\mathbbm1_{N_2}\oplus\,\ldots\oplus\,\bar\omega \mathbbm1_{N_{n-1}}.
\end{equation}
This action is more straightforward to see on the bosonic fields of the effective theory.
Open strings ending on the $N$ D branes in the unprojected theory give rise to a $U(N)$ super Yang--Mills theory. Focusing on D3 branes, one has an $\mathcal N=4,\ D=4$ effective theory. The bosonic matter content comprise one vector $A_\mu$ and three complex scalars $\varphi^I$ arising from the reduction of the ten-dimensional vector.
The orbifold projection will break supersymmetry to $\mathcal N=2$ or 1.
The vector in $D=4$ is not directly affected by the $\mathbb Z_n$ action on the internal space coordinates \eqref{orientifold action on internal coords},  but the action on the Chan--Paton indices remains, so that
\begin{equation}
	A_\mu\,{}_i{}^j \rightarrow  \left(\rho A_\mu \rho^{-1}\right)_i{}^j = \omega^{i-j} A_\mu\,_i{}^j.
\end{equation}
Here $U(N)$ indices are grouped in their conjugacy classes. We must only keep invariant states, hence we require $i=j$ and the gauge group is reduced to $\prod_i U(N_i)$.
Scalar fields arising from the decomposition of the $D=10$ vector transform under $\mathbb Z_n$ as the $z^I$. Combining with the Chan--Paton action one gets
\begin{equation}
 \varphi^I {}_i{}^j \rightarrow \omega^{k_I}  \left(\rho\, \varphi^I \rho^{-1}\right)_i{}^j = \omega^{i-j+k_I} \varphi^I{}_i{}^j
\end{equation}
Therefore the matter content surviving the projection sits in the bi-fundamental representations $\bf N_i\bar N_j$ of $\prod U(N_i)$, where $i-j+k_I = 0 \mod n$.
The resulting gauge theories can be conveniently depicted in a quiver diagram, where nodes represent gauge groups while arrows connecting nodes indicate matter in the bi-fundamental representation. Orbifold quiver theories have a tree-level super-potential for the chiral superfields $\Phi^I$ of the form $W_{\rm tree} = \Tr_N\Phi^1[\Phi^2,\Phi^3]$, that is inherited form the parent $\mathcal N=4$ super Yang--Mills theory.

One can combine orbifold projections with an un-oriented projection. Adding an ${\Omega}3$-plane on top of the `fractional' D3 branes, one identifies different groups and states. When $n$ is odd, the $ {\Omega}$ projecton must necessarily identify one of the nodes of the quiver with itself, hence the only consistent identification is 
\begin{equation}
 N_0 = \bar N_0,\quad N_i = \bar N_{n-i}
\end{equation}
The first condition gives either $SO(N_0)$ or $Sp(N_0)$ depending on the $\Omega$-plane charge.
For even $n$, there are two possibilities:
\begin{gather}
	N_0 = \bar N_0,\quad N_{n/2} = \bar N_{n/2},\quad N_i = \bar N_{n-i};  \\[.3em]
	N_0 = \bar N_{n/2}, \quad N_i = \bar N_{n/2-i}.
\end{gather}
Moreover there is the choice of the charge of the $\Omega$ plane, leading to different possibilities. One must also require RR tadpole cancellation in sectors with non-vanishing (worldsheet) Witten index, reflecting anomaly cancellation for the non-abelian gauge group in the effective field theory \cite{Anom&Tad, AIMU}. The situation can be summarized in terms of \eqref{orientifold action on CP group} with the requirement $\Tr \rho(\mathbb Z_n) = \pm q_{\Omega}$, where $q_\Omega$ is the charge of the orientifold plane, when present.

\begin{figure}
\label{fig:quivers}
\begin{center}
\includegraphics[scale=1]{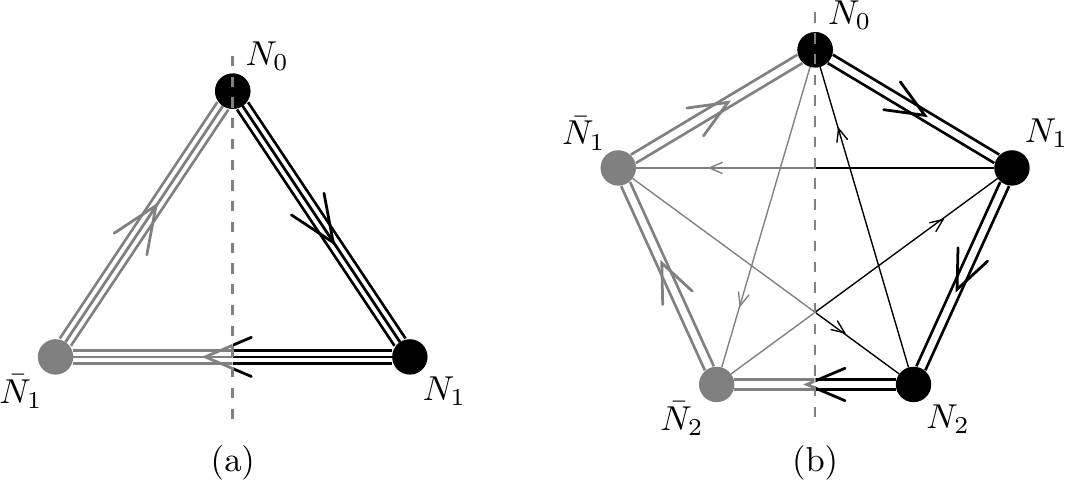}
\caption{%
	The quiver diagrams for (a) $\bbC^3/\bbZ_3$  and (b) $\bbC^3/\bbZ_5$.
	The dashed lines represent the unoriented projection.
}
\end{center}
\end{figure}

For instance, let us consider the reduction of $\mathcal N=1$ super Yang--Mills theory on $T^6/\mathbb Z_3$ \cite{ABPSS}.
Internal space coordinates are identified according to $z^I \simeq \omega z^I$, $\omega=e^{2\pi i/3}$ and the gauge group turns out to be
$U(N_0)\times U(N_1)\times U(N_2)$ with three generations of chiral multiplets in the
$\bf N_0 \bar N_1 \oplus N_1 \bar N_2 \oplus N_2 \bar N_0$.
The corresponding quiver diagram is shown in Figure \ref{fig:quivers}.
The unoriented projection is obtained with an ${\Omega}3^{\pm}$ plane, which leads either to a $SO(N_0)\times U(N_0+4)$ gauge theory (taking into account anomaly cancellation) or to $Sp(2N_0)\times U(2N_0-4)$, for negative or positive charge respectively.

Quiver gauge theories generally contain gauged $U(1)$ factors for which anomaly cancellation is not just obtained by the appropriate choice of the charges of matter fields, but rather it is recovered from the contribution of twisted RR axions $\zeta$ coming from closed string modes and generalized Chern-Simons couplings \cite{Pascal, Lionettoetal}.
For a single anomalous $U(1)$ associated to the gauging of an axionic shift symmetry, the tree-level lagrangian has the schematic form:
\begin{equation}
 \cL_{\rm ax} = (\partial_\mu \zeta - M A_\mu)^2 + C\, \zeta \Tr F\wedge F.
\end{equation}
Under a gauge transformation $\delta A_\mu = \partial_\mu \alpha ,\ \delta\zeta = M\alpha$, the second term is not invariant. However, it compensates the variation of the anomalous 1-loop contributions from the fermions, provided $M$ and $C$ are suitably chosen.

As we will momentarily see, closed string twisted modes actually enter the expression of non-perturbative super-potentials  in the form $\sim e^{-C Z}$, so that a shift of the axion $\zeta = \Im Z$ compensates the $U(1)$ charge of the factors depending on charged open string fields.

\section{Non-perturbative superpotentials in the $T^6/\bbZ_3$ models}

We provide here a more detailed discussion of the unoriented projections of the  $T^6/\bbZ_3$ quiver gauge theories \cite{U(4)}.
In particular, we would like to discuss how non-perturbative superpotentials arise from gauge and exotic instantons in the $SO(N_0)\times U(N_0+4)$ and $Sp(N_1+4)\times U(N_1)$ models.
As already explained, they arise as the $\cN=1$ effective theories of stacks of D3 branes on the $T^6/\bbZ_3$ orbifold singularity combined with an ${\Omega}3^{\pm}$ plane.
Gauge and exotic instantons are obtained by considering bound states with D$(-1)$ and ED3 branes (wrapping 4-cycles in the internal space) respectively.

\subsection{Gauge instantons}

The construction of gauge instantons is obtained from the standard D$(-1)$ D3 system after implementing both unoriented and orbifold projections.
We begin by considering the unoriented projection, which sends the $U(K)$ instanton symmetry either to $SO(K)$ or $Sp(K)$. Consistency requires the D$(-1)$ D$(-1)$ system to obey the opposite identifications with respect to the D3 system, so that the relevant symmetry groups are $SO(N)\times Sp(K)$ and $Sp(N)\times SO(K)$ for negative and positive ${\Omega}$3 charge respectively.
The $\bbZ_3$ identification splits the $N$ D3 and $K$ D$(-1)$ branes in $N_0,\ N_1$ and $K_0,\ K_1$ fractional branes. The relevant instanton symmetry group becomes either $Sp(K_0)\times U(K_1)$ or $SO(K_0)\times U(K_1)$.
The surviving ADHM data is obtained following the same lines as in Section 3 and keeping states invariant under $\bbZ_3$ and the unoriented projection.
The resulting representations for the ${\Omega}3^+$ case are:
\begin{align*}
 a_\mu ,\ \Theta^0_\alpha:			&\qquad {\mathbf{K_0}\otimes_s \mathbf{K_0}}\ \oplus\ {\rm Adj}(U(K_1))	%
 & \Theta^I_\alpha:						&\qquad {\mathbf{K_1}\otimes_s \mathbf{K_1}}\ \oplus\ \bf K_0\bar K_1		
 \\
 D^q,\ \bar\Theta_{0\,\dot\alpha}: &\qquad {\rm Adj}(SO(K_0)) \ \oplus\ {\rm Adj}(U(K_1))%
 &\bar\chi_I,\ \bar\Theta_{I\,\dot\alpha} : &\qquad {\mathbf{\bar K_1}\otimes_a \mathbf{\bar K_1}}\ \oplus\ \bf K_0\bar K_1%
 \\	
 \chi^I: &\qquad  {\mathbf{K_1}\otimes_a \mathbf{K_1}}\ \oplus\ \bf K_0\bar K_1%
& w^{\dot\alpha},\ \nu^0: &\qquad \bf K_0N_0 \oplus K_1\bar N_1 \oplus N_1\bar K_1%
 \\
 \nu^I: &\qquad \bf K_0\bar N_1 \oplus \bar K_1 N_0 \oplus K_1N_1%
\end{align*}
We decomposed $A=0,I$ where $A=0$ indicates the surviving Killing spinor direction.
Of course, no independent $\bar w, \bar\nu$ moduli are present in the unoriented theory. The content for $\Omega3^-$ is obtained by exchanging $\otimes_s \leftrightarrow \otimes_a$ and $SO(K_0) \leftrightarrow Sp(K_0)$.

It can be shown that the only configurations which give rise to a non-perturbative superpotential are either the $U(4)$ model (with an $ {\Omega}3^-$ plane) with $K_0=0,\ K_1=1$ or the $Sp(6)\times U(2)$ theory with $K_0=1,\ K_1=0$. This is again obtained by counting of the fermionic zero-modes.

In the $Sp(6)\times U(2)$ case the relevant gauge instanton would have $O(1)$ symmetry and therefore most moduli disappear: the whole ADHM data, with explicit indices  for the gauge group, is
\begin{equation}
 a_\mu ,\ w^{\dot\alpha}_s,\ \Theta^0_\alpha ,\ \nu^{0\,s},\ \nu^{I\,u},\qquad s=1,\ldots,6;\quad u=1,2.
\end{equation}
In particular no $\bar\Theta$ and $D^q$ are present and therefore no ADHM constraint must be imposed. Only two fermionic zero-modes $\Theta^0_\alpha$ survive, as required.
The instanton action reduces to the couplings $S_\varphi$ of charged moduli to dynamical scalars $\varphi^{Isu}$, as in \eqref{S phi}.
Integrations along $w,\ \nu$ are gaussian and hence \eqref{ADS} yields the final result
(substituting $a_\mu\rightarrow x_\mu$, $\Theta^0\rightarrow\theta$ up to a nonvanishing constant):
\begin{equation}
 W_{np,Sp(6)} = \int d^4\hspace{-1pt}x\ d^2\hspace{-1pt}\theta \frac{\Lambda^9}{\det_{6\times6}(\Phi^{Iu,s})}.
\end{equation}
The exponent of $\Lambda$ is given by the one-loop coefficient of the $\beta$ function or, conversely, can be fixed a posteriori by dimensional analysis. We treat $Iu$ as a single index from 1 to 6.

The case of the $U(4)$ theory is slightly more involved as some more moduli survive and there is an actual $U(1)$ instanton symmetry. We give only the result, in terms of the chiral superfields $\Phi^{I[uv]}$ in the antisymmetric representation of the gauge group ($u,v=1,\ldots,4$).
\begin{equation}
W_{np,U(4)}  = \int d^4\hspace{-1pt}x\ d^2\hspace{-1pt}\theta \frac{\Lambda^9}{\det_{3\times3}(\Phi^{Iuv}\Phi^{Jwt}\epsilon_{uvwt})}.
\end{equation}

\subsection{Exotic instantons}

Exotic instantons in the $T^6/\bbZ_3$ model are generated by (fractional) ED3 branes wrapping a 4-cycle in the internal space \cite{U(4)}.
Open strings connecting ED3 and D3 branes have 8 mixed Neumann--Dirichlet directions. Thus the ground-state in this sector is a fermion of fixed chirality (say R) with respect to the $SO(2)$ acting on the common transverse directions \cite{PWDH, EW}.

A counting of the allowed exact fermionic zero-modes shows that a superpotential contribution can only arise for the $SO(N_0)\times U(N_1)$ case, with $K_0=1,\ K_1=0$. Note that this time the symmetry group of the instanton, arising from the stacks of fractional ED3 branes, is of the same kind of the gauge group, hence $SO(K_0)\times U(K_1)$, as a consequence of the unoriented projection acts in the same way on the physical and instantonic branes.

The superpotential calculation is similar to the gauge instanton case, but this time the instanton has no sizes and orientations and hence the integration over charged moduli yields positive powers of the fields:
\begin{equation}
 W_{ex,U(4)}  = \Lambda^{-\frac{N_1}{2}+3} \int d^4x\ d^2\theta d^{N_1} \nu\  \exp(-\nu_u \varphi^{uv} \nu_v).
\end{equation}
The scale factor is calculated in a similar way from tree- and one-loop amplitudes, however it is not directly related to the gauge coupling constant $\tau$.
The resulting superpotentials carry $N_1/2$ power of the dynamical fields and therefore are renormalizable for $N_1\leq6$.

For instance, for the $U(4)$ model one obtains non-perturbative Majorana mass terms for the chiral fields in the three ${\bf 6}$, while for the $SO(2)\times U(6)$ with three $({\bf 2},{\bf 6}_{-1}^*) + ({\bf 1},{\bf 15}_{+2})$, one finds instead corrections to the Yukawa couplings of the ${\bf 15}_{+2}$ with one another.
Notice that in both cases, exotic instantons generate couplings for the matter in the antisymmetric representation.

\section{The $\bbZ_5$ quiver}

Another interesting model is obtained from (unoriented) D3 branes at a $\bbC^3/\bbZ_5$ singularity \cite{MBJFMFF}. Note that although the torus $T^6$ does not admit a $\bbZ_5$ projection compatible with supersymmetry, such singularities can show up in Calabi--Yau compactifications.
The quiver diagram is depicted in Figure \ref{fig:quivers}. After the unoriented projection and keeping into account tadpole cancellation, one can construct in particular a GUT-like model with $U(5) \times U(1)$ gauge group which features dynamical supersymmetry breaking.
\vspace*{.3cm}
\begin{vchtable}[ht]
\begin{center}
\begin{tabular}{|c|c|c|c|}
\hline
fields & $SU(2)_{F}$ & $SU(5)$ & $U(1)_1\times U(1)_5$ \\
\hline
$A^{i  }  _{uv}$ & ${\bf 2}$  & ${\bf 10}$ & $(0,2)$ \\
$B^{i  }  $ & ${\bf 2}$  & ${\bf 1}$ & $(-1,0)$ \\
$C^{{i  }   u} $ & ${\bf 2}$  & ${\bf \bar 5}$ & $(1,-1)$ \\
  $C^{3 u}$ & ${\bf 1}$  & ${\bf \bar 5}$ & $(-1,-1)$ \\
   $E_{u}$ & ${\bf 1}$  & ${\bf  5}$ & $(0,1)$ \\
\hline
\end{tabular}
\end{center}
\caption{\small Chiral matter field content. Indices ${i  }  =1,2$, $u=1,..5$ run over the fundamentals of the $SU(2)$
flavor and $SU(5)$ gauge
groups. }
\label{matter}
\end{vchtable}%

The matter content of the $U(5) \times U(1)$ model is summarized in Table \ref{matter}.  Gauge kinetic functions are linear combinations of the two (closed string) twist fields $T_a= t_a + i \zeta_a$ and the axio-dilaton $S=\varphi+i b$. F-terms are easily obtained from the tree-level super potential
\begin{equation}\label{Wtree Z5}
W_{\rm tree} = C^{iu} B_i E_u + C^{iu} A_{iuv} C^{3v}.
\end{equation}

A counting of the fermionic zero-modes for this theory shows that no gauge instanton corrections to the superpotential can arise. However, such effects can appear in a less direct way, when the theory is in an appropriate regime.
Note moreover that no exotic instanton effects do affect the superpotential  \cite{MBJFMFF}.

Including Fayet--Iliopoulos terms, the D-terms read
\begin{equation}
 \begin{split}
 (D_{U(5)})^u{}_v &= \bar A_i{}^{uw} A^i{}_{wv} - \bar C_{iv} C^{iu} - \bar C_{3v} C^{3u} + \bar E^u E_v +\xi_{U(1)_5} \delta^u_v  \\
 D_{U(1)}\ \  &= -\abs{B_i}^2 +\abs{C^{iu}}^2 - \abs{C^{3u}}^2 + \xi_{U(1)}.
 \end{split}
\end{equation}
For $\xi_{U(1)_5} = 0,\ \xi_{U(1)} = m^2 >0$ there is an $SU(5) $ preserving solution to both F- and D-flatness conditions, which defines a perturbative supersymmetric vacuum of the theory, obtained by taking all vev's to vanish except for the $U(5) $ singlet:
\begin{equation}
 \langle B_i \rangle = (m,\, 0).
\end{equation}
The superpotential produces a mass term $ W_{\rm tree} = m C^{1u} E_v $ for $C^{1u}$ and $E_u$,
that therefore decouple for large $m$. The low energy tree-level superpotential reads $W = A^1_{uv} C^{2u} C^{3v}$.
The resulting theory is a $U(5)$ gauge theory with 2 generations of matter in the $\bf 10$ and $\bf\bar 5$ representations (plus two color singlets).
What is interesting is that now an effective superpotential can arise from gauge instantons.
Indeed, the counting of fermionic zero-modes from a $K=1$ $U(5)$ instanton is $n(\lambda_0) - n(\psi_0) = 2N - \sum_{\bf R} \ell_2 ({\bf R}) = 10 - (1+3)\times2 = 2$.
The exact form of the gauge instanton correction to the superpotential is also completely fixed by the symmetries of the theory
\begin{equation}\label{ADS Z5 superpotential}
 W_{ADS} = \frac{\hat \Lambda^{11}}{Z^{i\alpha}Z_{i\alpha}}, \qquad Z^{i\alpha} = \frac{1}{12}\epsilon^{u_1\ldots u_5} A^i_{u_1 u_2} A^j_{u_3 u_4} A_j{}_{u_5 v}C^{\alpha v}.
\end{equation}
Here the matching of the scales of the two theories gives $\hat \Lambda^{11} = m\Lambda^{10}$.
The new index $\alpha$ runs over the 2 remaining massless combinations of the $C$ fields.
Taking into account this non-perturbative correction, the effective $U(5)$ theory no longer admits a solution to both F- and D-flatness conditions, so that supersymmetry is dynamically broken by the instanton effect.

It is very interesting to see how one can recover the same results in the strong coupling description  \cite{MBJFMFF}. In this case, the appropriate degrees of freedom are  $SU(5)$ ``color'' singlets: the $U(1)$ factors are anomalous and at any rate decouple at low energy.
Introducing an index $I=(i,3)$, one can construct ``mesons'' and ``baryons'':
\begin{equation}\label{mesons and baryons}
\begin{split}
 X^I 		&= C^{Iu} E_u,\quad&
 Y^i_I 	&= \frac{1}{2}\epsilon_{IJK} A^{i}_{u_1 u_2} C^{J u_1} C^{K u_2}, \\
 \tilde Y^{ij} &= \frac{1}{4}\epsilon^{u_1\ldots u_5} A^i_{u_1 u_2} A^i_{u_3 u_4} E_{u_5}, \quad&
 Z^{iI} &= \frac{1}{12}\epsilon^{u_1\ldots u_5} A^i_{u_1 u_2} A^j_{u_3 u_4} A_j{}_{u_5 v}C^{I v}. \\
\end{split}
\end{equation}
Note that now the flavour symmetry is apparently extended to $SU(3)\times SU(2)$, although only a diagonal $SU(2)$ is preserved by the superpotential.
The fields in \eqref{mesons and baryons} satisfy the algebraic relations
\begin{gather}\label{tree level constraints}
 Y^i_I Z^I_i = 0,\qquad
 \epsilon_{IJK} X^I Z^J_i Z^{iK} + Y_{iI} \tilde Y^{ij} Z^I_j = \Lambda^{10}.
\end{gather}
The term $\Lambda^{\beta_1}=\Lambda^{10}$ arises at the quantum level and is absent at tree level. It represents a quantum deformation of the moduli space. Note that it could not appear in the first relation since the terms there have the `wrong' dimension: 7 instead of 10.
One can rewrite the superpotential \eqref{Wtree Z5} by adding Lagrange multipliers $U$ and $V$ to enforce the constraints
\begin{equation}\label{Weff strong}
 W_{\rm eff} = B_i X^i + \delta^I_i Y^i_I + V Y^i_I Z^I_i + U ( \epsilon_{IJK} X^I Z^J_i Z^{iK} + Y_{iI} \tilde Y^{ij} Z^I_j - \Lambda^{10} )
\end{equation}
Turning on a vev for the  singlet field $B^1$and integrating out the massive fields, one is left with $Z^{i\alpha}$ and
$Y^i_1$, together with the color singlets $B_i$. They match the number of degrees of freedom of the weak coupling description, namely $32 -24= 8$, where subtraction is due to $SU(5)$ gauge invariance.
The resulting superpotential takes the form $W_{\rm eff} = Y^1_1 + \frac{m \Lambda^{10}}{Z^{i\alpha}Z_{i\alpha}}$, hence matching the ADS correction of the effective theory. Again, we conclude that supersymmetry is dynamically broken.

Another way to see this result relies on the Konishi anomaly. One starts by considering a correlator in the microscopic weakly coupled theory that is corrected by instanton effects. In the case at hand, it has the schematic form $\langle E C^3 A^6 \rangle$, since the ten scalars soak up two fermionic zero modes each and hence saturate the counting of 20 fermionic zero-modes of the full theory for a $K=1$ instanton. Written in terms of \eqref{mesons and baryons} for gauge invariance, one has
\begin{equation}
 \langle X^I Z^{iJ} Z^{jK} \rangle = \epsilon^{IJK} \epsilon^{ij} \Lambda^{10}.
\end{equation}
Note that the correlator is independent of the insertion points and hence effectively factorizes. One can then use the Konishi anomaly relations \cite{Amati:1988ft,Bianchi:2007ft}:
\begin{equation}\label{konishi anomaly}
 \frac{1}{2\sqrt2}\langle \{\bar Q_{\dot\alpha},\, \bar\psi_{\dot\alpha}^f  \phi_{f'}\} \rangle
  = \langle\phi^f \frac{\partial W_{\rm tree}}{\partial \phi^{f'}}\rangle + \frac{g^2}{32\pi^2} \delta^f{}_{f'} \Tr \langle \lambda\lambda\rangle.
\end{equation}
By taking $f=f'$ and $\phi^f= E$ and the vev's $\langle B_i \rangle= (m,0)$ and noting that the left hand side of \eqref{konishi anomaly} vanishes unless supersymmetry is broken and $\psi^f$ is the Goldstino, we have
\begin{equation}
 \frac{g^2}{32\pi^2}\langle \Tr\lambda\lambda\; Z^{i\alpha}Z_{i\alpha} \rangle = m\Lambda^{10}
\end{equation}
which guarantees the presence of a gaugino condensate $\Tr \langle \lambda\lambda \rangle \neq 0$. Now, taking \eqref{konishi anomaly} with $\phi^f=\phi^{f'}=A^2$ we see that $\langle\phi^f \frac{\partial W_{\rm tree}}{\partial \phi^f}\rangle =0 $ in the low energy theory where massive fields have been integrated out, which means that the left hand side must be nonvanishing. Therefore supersymmetry is dynamically broken and we identify the Goldstino with $\psi_{A^2}$.

\section{Outlook}

Let us conclude with an outlook and draw some directions for future work
on the subject.

We hope we have convinced the reader that D-brane instantons (ED-branes)
can produce interesting non-perturbative effects in vacuum configurations
with open and unoriented strings. As we have shown, there are two broad
classes of ED-branes: `gauge' instantons and `exotic' instantons. The
former provide a very elegant way of embedding the ADHM construction in
string theory. The latter are genuinely stringy in nature. Depending on
the amount of supersymmetry and the structure of fermionic zero-modes,
ED-branes can produce non-perturbative super-potentials, threshold
corrections to gauge kinetic functions or to higher-derivative terms, etc.
They can in particular contribute to moduli stabilization and
supersymmetry breaking in conjunction with internal fluxes and FI D-terms.

We have neither explored the interplay with world-sheet and NS5-brane
instantons, resulting from various string dualities, nor the holographic
description in various contexts that could shed light on multi-instanton
effects and other subtle properties of ED-branes, such as the role of
world-volume fluxes. Although there is still a long way to go, we believe
the  basic rules of the  game are set for future investigation.

\begin{acknowledgement}
We would like to thank the organizers for the kind invitation. We would
like to thank L. Martucci, D. Ricci Pacifici, R. Richter and Ya. Stanev
for discussions. M. B. would like to thank F. Fucito, E. Kiritsis, S.
Kovacs, J. F. Morales, R. Poghossian, G. Pradisi, G.C. Rossi for
collaboration on topics related to this lecture. This work is partly
supported by Italian MIUR-PRIN 2009KHZKRX-007 "Simmetrie dell'Universo e
delle Interazioni Fondamentali", by the ERC Advanced Grant No 226455
``Superfields" and by the ESF Network ``HoloGrav".
\end{acknowledgement}


\end{document}